\documentclass[aps,pra,twocolumn,showpacs,preprintnumbers,amsmath,amssymb,footinbib]{revtex4}
\usepackage{graphicx,epsfig}
\usepackage{placeins}
\usepackage{bm}
\usepackage{dcolumn}
\usepackage{xcolor}
\usepackage[breaklinks=true,colorlinks,citecolor=blue,linkcolor=blue,urlcolor=blue]{hyperref}
\usepackage{xr}
\usepackage{siunitx}

\def\noi{\noindent}
\def\bc{\begin{center}}
\def\ec{\end{center}}
\newcommand{\bea}{\begin{equation}}
\newcommand{\eea}{\end{equation}\noi}
\newcommand{\ber}{\begin{eqnarray}}
\newcommand{\eer}{\end{eqnarray}\noi}
\begin{document}
\title{Reconfigurable multiplex setup for high throughput electrical characterisation at cryogenic temperature}

\author{X Bian$^{1}$}\email{xinya.bian@materials.ox.ac.uk}
\author{H J Joyce$^2$, C G Smith$^3$, M J Kelly$^2$, G A D Briggs$^1$}
\author{J A Mol$^{4}$} \email{j.mol@qmul.ac.uk}
\affiliation{$^{1}$Department of Materials, University of Oxford, Oxford, OX1 3PH, UK\\
$^{2}$Centre for Advanced Photonics and Electronics, Department of Engineering, University of Cambridge, CB3 0FA, UK\\$^{3}$ Department of Physics, Cavendish Laboratory, University of Cambridge, CB3 0HE, UK \\$^{4}$ School of Physics and Astronomy, Queen Mary University, London, E1 4NS, UK}
\date{\today}

\begin{abstract}
In this paper, we present a reconfigurable multiplex (MUX) setup that increases the throughput of electrical characterisation at cryogenic temperature. The setup separates the MUX circuitry from quantum device under test (qDUT), allowing qDUT chips to be exchanged easily and MUX chips to be reused. To interface with different types of qDUTs, board-level designs are incorporated to allow interconnects flexibly routed into different topology. MUXs are built based on a multiple level selective gating (MLSG) scheme, where the number of multiplexed output channels (interconnects) is exponentially dependent on the number of control lines. In the prototype setup presented in this paper, with 14 out of 44 existing wires from room temperature, 4 MUXs at cryogenic temperature can supply in total 128 interconnects to interface with qDUTs. We validate the MUX setup operation and assess the various limits existed by measuring k$\Omega$ resistors made of $\mu$m-size graphene ribbons. We further demonstrate the setup by performing charge transport measurement on 128 nm-size graphene quantum devices in a single cooling down.
\end{abstract}

\maketitle

\section{Introduction}
To develop new technologies that exceed the limits of classic electronics, different types of quantum devices have been proposed and demonstrated \citep{Zwanenburg2013,Sarma2015,Watson2018,Xiang2016}. However, most candidates are still limited to small scale circuit operation and lack a viable way to scale up. A bottleneck is the limited throughput of electrical characterisation at cryogenic temperature, which slows down the optimisation process towards high reproducibility. To overcome the problem requires large numbers of qDUTs to be tested in a single cooling down and constraints from two aspects need to be dealt with. One is the various constraints (e.g. cooling power, physical space) imposed by operating at cryogenic temperature, particularly the limited number of wires allowed from room temperature. The other is the inherent complexity to interface with qDUTs. To operate a qDUT typically requires multiple signal lines with individual tunability. For example, 8 signal lines are needed to operate an electrically defined double quantum dot, while only a few tens of wires are available in a typical crystat setup, thus only a few devices can be tested in a single cooling down. 

There have been some attempts made towards addressing this throughput issue so far. One approach is to build cryogenic wafer scale probe station. As reported recently, quantum dot type devices were characterised at a wafer temperature as low as 1.9K, which however only allows certain device characteristics to be measured \citep{Pillarisetty2019}. Other key metrics requiring lower temperature and high magnetic field still rely on measurement inside mK dilution refrigerators, of which the throughput is still limited. An alternative approach is to incorporate the MUX circuitry to increase the number of interconnects at cryogenic temperature, which has the advantage to be compatible with mK dilution refrigerator. Two initial works have the MUX circuitry integrated on chip \citep{Al-Taie2013,Ward2013}, which makes them less suitable for fast prototype, test and optimisation. Besides the additional complexity involved in fabrication, changes in qUDT design are difficult to be accommodated. For instance, adding an extra gate electrode in quantum dot device to define a quantum point contact for probing charge occupancy will require a completely new design. Two more recent solutions \citep{Wuetz2020,Pauka2019} have implemented the MUX circuitry off-chip instead. However, they are still with their own drawbacks. Both solutions are still with limited flexibility to alter the interconnect topology to meet different wiring requirements. Secondly, both off-chip solutions rely on a digital clock driven serial in parallel out (SIPO) shift register to control MUXs. As discussed by themselves, the complexity arises naturally from the requirement to match digital signal delay at cryogenic temperature and to decouple digital noise. 

In this paper, we present a cryogenic MUX setup that can be easily configured to meet different wiring requirements. The MUXs are built based on the MLSG method as previously used in \citep{Al-Taie2013,Ward2013}, which does not rely on a digital clock signal to function. We implement the MUXs off-chip, easing the exchange of qDUT chips and reusing the MUX chips. To accommodate changes made to qDUTs, we also incorporate schemes to allow interconnects to be routed into different topology at the qDUT side. Moreover, the MUX operation is supported by a shared control line arrangement. Such arrangement not only facilitates efficient scaling up the total number of interconnects, but also simplifies the simultaneous control of multiple independent signals supplied from different MUXs. As a result, qDUTs that require multiple signal lines to operate can be interfaced easily. We first validate and assess the operation of MUX setup by measuring $\mu$m-size graphene ribbons. We further demonstrate the setup by performing transport measurement on 128 nm-size graphene quantum devices. Lastly, we discuss the potential improvements that can further increase the number of interconnects and estimate that it can supply more than 1000 interconnects if all the improvements are implemented.

\section{Multiplex measurement setup details}
\begin{figure*}[h]
\includegraphics[width=0.98 \linewidth]{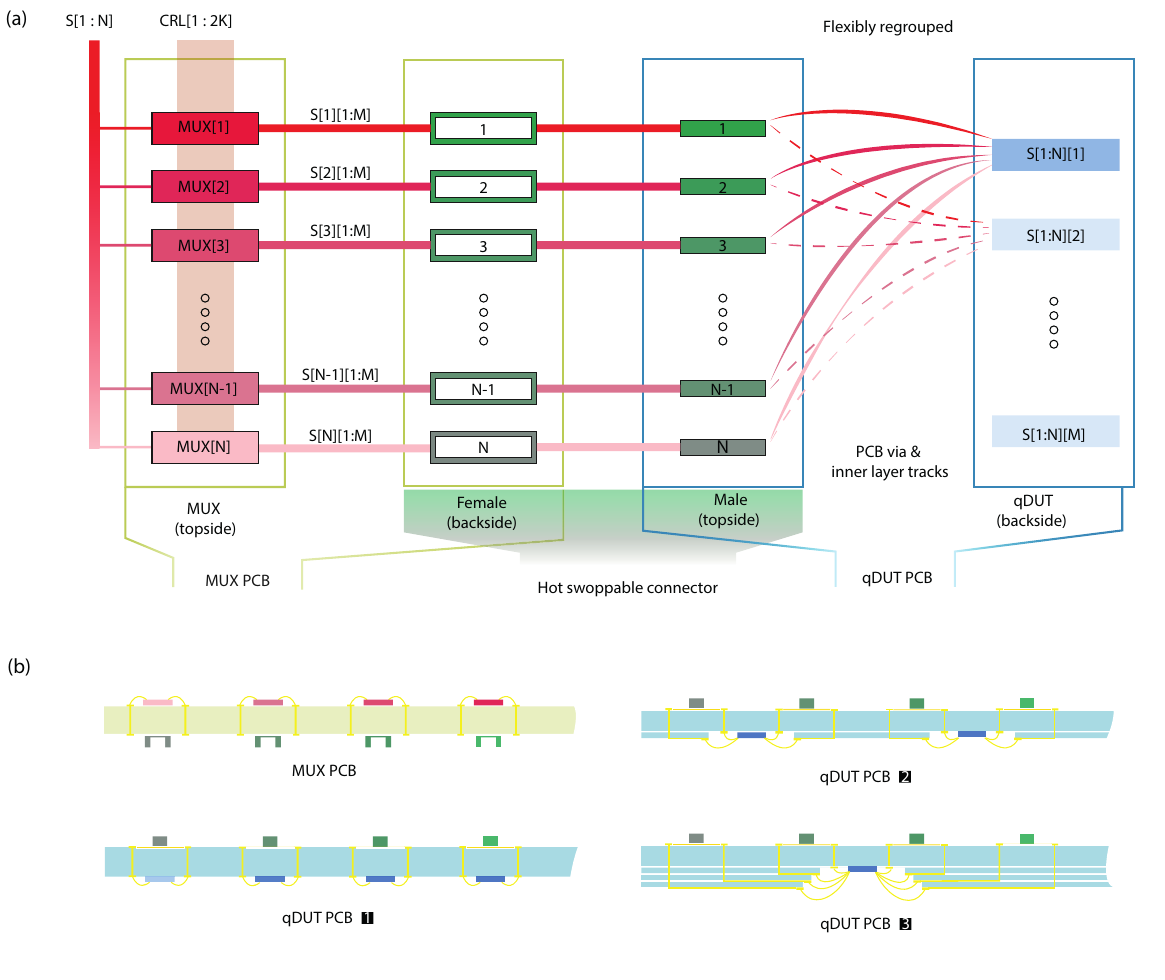}
\caption{(a) Schematics of the MUX setup. The setup comprises two PCBs, of which one is for mounting MUX chips and the other is for mounting qDUT chips. Two PCBs are connected via swoppable board to board connectors. The interconnects can be regrouped differently at the qDUT PCB side to meet different wiring requirements. Each MUX (MUX[1], MUX[2], ..., MUX[N]) is supplied with a input signal (S[1], S[2], ..., S[N]) from room temperature and converts it to M individually addressable outputs. MUXs are controlled by the shared control lines CRL[1:2K]. With 2K control lines, each MUX can supply up to 2\textsuperscript{K} output channels, i.e. M = 2\textsuperscript{K}. (b) Cross section view of the MUX setup. MUX chips are wire bonded to the top side of MUX PCB. Interconnects supplied from MUX chips are connected to the female connectors at the backside of MUX PCB through the PCB vias. qDUT chips are wire bonded to the backside of qDUT PCB, which are then connected through PCB vias with the male connectors at the top side of qDUT PCBs. The MUX PCB stay unchanged once being installed and different qDUT PCBs can be used for mounting different qDUTs. qDUT PCB 1, 2, 3 can be used to interface with qDUTs that require 2, 4, 8 signals to operate respectively. The multiple layer PCBs used here allows multiple tier wire bonding to be used.
\label{Schematics}}
\end{figure*}

Figure \ref{Schematics}a shows the schematics of the MUX setup and gives an overview of how signals are multiplexed and routed at cryogenic temperature. The setup consists of two print circuit boards (PCBs), of which one is for mounting MUX chips and the other is for mounting qDUT chips. The MUX PCB and qDUT PCB are connected via swoppable board to board connectors. The MUX PCB comprises N MUXs and each MUX is connected to a input signal line from room temperature converting it to M individually addressable output channels. The MLSG method \citep{Al-Taie2013,Ward2013} used to implement signal multiplexing ensures great scalability, as the number of multiplexed output channels grows exponentially with the number of control lines. For a K-level base-2 MUX, 2K control lines are needed for addressing gates and the multiplex ratio M is equal to 2\textsuperscript{K}. Physically, base-2 MUX corresponds to each channel at current level being split into two channels at next level, thus leading to the exponential increase. Control lines are supplied to the MUX PCB and shared among all the MUX chips, adding another layer of scalability. Thus, N~+~2K wires from room temperature are able to supply in total N $\times$ 2\textsuperscript{K} interconnects at cryogenic temperature. To scale up the total number of interconnects, it can be accomplished either by using MUXs of more levels or by increasing the number of MUXs. For every extra level added to MUXs, two more control lines are required and the total number of interconnects doubles. For every K-level base-2 MUX added to the setup, only one extra signal line is required and the total number of interconnects increases by 2\textsuperscript{K}.

As the control lines are shared among MUXs, the same multiplexed output channel of different MUXs can be simultaneously selected and set to supply completely independent signals. Such a shared control line arrangement provides a very simple way to control multiple MUXs to interface with qDUTs. For a setup with N MUXs, it is able to interface with qDUTs requiring up to N independent signals to operate. Certainly, qDUTs that need fewer independent signals to operate can also be interfaced. The reconfiguration of the setup to meet different wiring requirements is straightforward and can be accomplished solely at the qDUT side. Figure \ref{Schematics}b shows how the interconnects supplied from the MUX PCB are routed into different topology in different qDUT PCBs. The qDUT PCB 1, 2 and 3 can be used to interface with qDUTs that require 2, 4, and 8 signal lines to operate respectively. Interconnects supplied from different MUXs are routed via different inner layer tracks, allowing multiple tier wire bonding to be implemented.

\begin{figure*}[h]
\includegraphics[width=0.98 \linewidth]{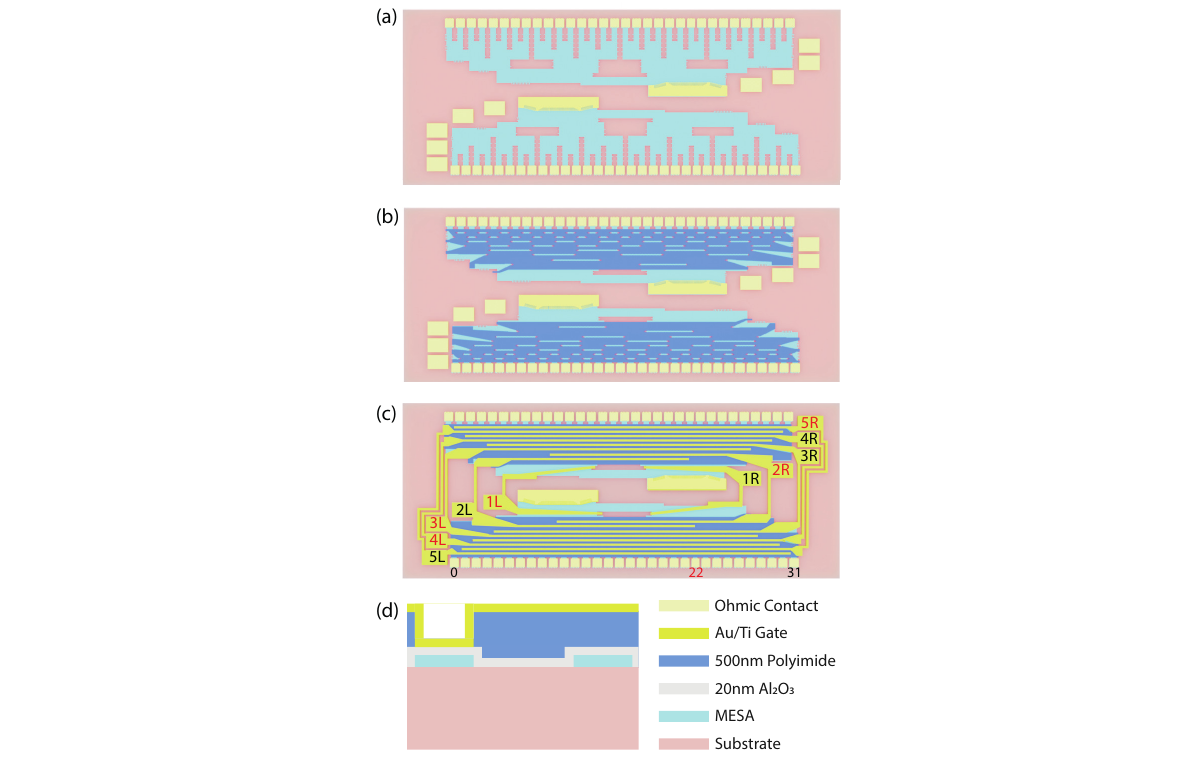}
\caption{MUXs are fabricated on GaAs substrate and each chip consists of two MUXs. (a) The conduction channels (MESAs) are first defined by wet etching, then Ohmic contacts are made for the input and each output. The isolated Ohmic contacts (and MESAs underneath) are used as bond pads for addressing gates. (b) 20~nm Al\textsubscript{2}O\textsubscript{3} is deposited globally (only shown in cross section view) and is etched to expose Ohmic contacts. Then polyimide is selectively deposited over the MESAs, which is used for modulating gate capacitance, such that channels can be selectively pinched off. (c) Au/Ti metal stack is deposited as addressing gates. By activating addressing gates 1L/2R/3L/4L/5R, channel 22 is selected. The relationship is clearly revealed in binary representation, if L corresponds to 1 and R corresponds to 0, then 10110 in binary is essentially 22 in decimal. (d) Cross section view of the MUXs showing two adjacent conduction channels, of which the gate capacitance is modulated by selectively deposited polyimide.
\label{MUX_32}}
\end{figure*}

For the prototype demonstrated in this paper, 5-level base-2 MUXs with a multiplex ratio of 32 = 2\textsuperscript{5} are used. With 4 MUXs installed, 128 interconnects are available at cryogenic temperature and only 14 wires (10 for addressing gates and 4 for inputs) from room temperature are used. Figure \ref{MUX_32} shows the schematics of the 5-level MUXs, each chip consists of two MUXs sharing 10 addressing gates. As compared with the MUXs demonstrated previously in \citep{Al-Taie2013}, an extra layer of 20~nm Al\textsubscript{2}O\textsubscript{3} is added in our MUXs (see Figure \ref{MUX_32}d). This extra layer of dielectric serves two purposes here. One is to improve the GaAs surface quality by a self-cleaning reaction \citep{Hinkle2008}, so to minimise the variation in MUX characteristics induced by surface trapped charges. More importantly, it will allow us to operate the MUXs over a extended signal range as compared with the original demonstration \citep{Al-Taie2013}. The signal range would otherwise be limited by the Schottky junction formed between Au/Ti addressing gates and MUX conduction channels.

The fabrication procedure of MUXs is as follows. The MUXs are fabricated on GaAs/AlGaAs substrate. Firstly, conduction channels (MESA) are defined by standard wet etching process and Ohmic contacts are made for the input and each output (Figure \ref{MUX_32}a). Secondly, 20~nm Al\textsubscript{2}O\textsubscript{3} is deposited globally by atomic layer deposition (ALD) process (only shown in cross section view Figure \ref{MUX_32}d) and windows are opened over Ohmic contacts by hydrofluoric acid etching. Then, photo definable polyimide is selectively deposited on top of the 20~nm Al\textsubscript{2}O\textsubscript{3} layer to modulate the gate capacitance (Figure \ref{MUX_32}b). Lastly, 10~nm Ti and 150~nm Au are deposited by tilted rotatory thermal evaporation to ensure continuous metalisation over the thick polyimide layer.

At cryogenic temperature, current conduction along channels is through the 2-dimensional electron gas (2DEG) formed at the interface of AlGaAs/GaAs. Channel conduction is normally on and can be pinched off by applying a negative voltage to addressing gates. Figure \ref{MUX_Operation}a shows the channel pinch off process as the 2DEG is gradually depleted. As the potential difference exceeds certain threshold $V_{T}$, the 2DEG underneath the gate becomes completely depleted and conduction is lost. By modulating the gate capacitance with selectively deposited polyimide, addressing gates will only affect channels that are not covered by polyimide, if the gate voltage applied is within a certain window (i.e. selective gating). As can be seen from Figure \ref{MUX_32}b, a pair of addressing gates at each level are designed to be complementary, of which one controls the odd channels and the other controls even channels. Therefore, each output channel can be individually selected by a specific gate combination. For example, output channel 22 can be selected by activating (applying pinch off voltage) addressing gate 1L/2R/3L/4L/5R (Figure \ref{MUX_32}c).

Below we validate the proper operation of this 5-level base-2 MUX, as well as assess the various limits that exist. For the MUX to operate properly, the selected (open) channel should present as a low-resistance path over the signal range of interest and the leakage current of all other unselected (closed) channels should be negligible. As will be shown below, normally the resistance of the selected channel is much smaller as compared with that of qDUT, however it can become very large even dominant when the signal applied to MUX input is above certain positive threshold, limiting the input signal range at the positive side. As for the negative side, it is to ensure the potential of active addressing gates stay sufficiently negative with respect to the conduction channels. The potential difference has to exceed the threshold voltage $V_{T}$ to maintain unselected channels at pinched off state. For a specific gate voltage applied, the input signal range at the negative side is thus limited. These two limits are not fundamental but need to be dealt with carefully. In contrast, the number of output channels in a single MUX is more of a fundamental limit. To ensure the leakage current being negligible, it is required to have the effective resistance of all closed channels in parallel to be much larger as compared with that of a qDUT. If the off-resistance of each closed channel is $R_{off}$ and the maximum resistance of a qDUT (resistance of a qDUT is usually gate and bias dependent) is $R_{max}$, then the total number of channels M is limited to be $\sim$ $ \frac{R_{off}}{10R_{max}}$ in practice.

\begin{figure*}[h]
\includegraphics[width=0.98 \linewidth]{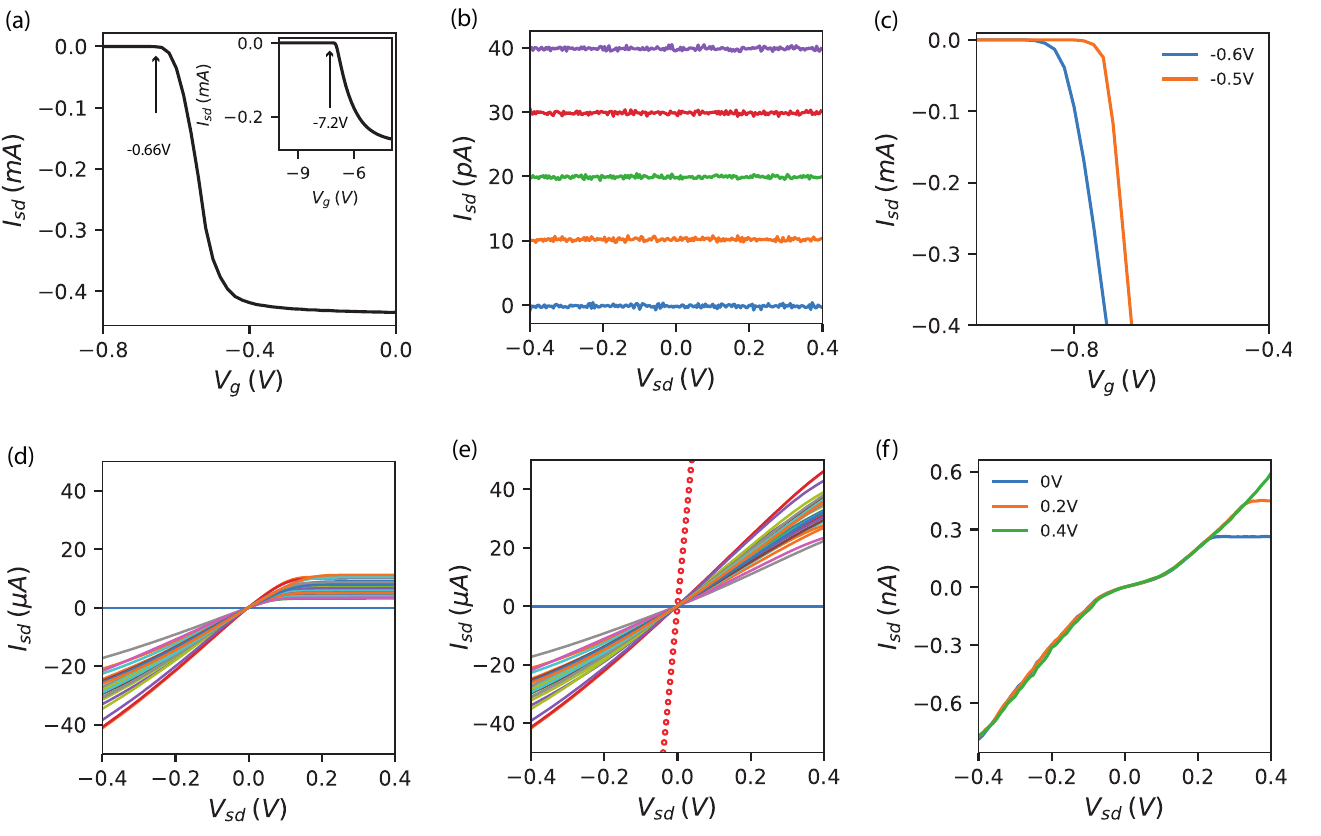}
\caption{(a) Pinch off characteristics of MUX channels that are not covered by 500~nm polyimide. It was measured with a input signal of $-0.4$~V and a pair of addressing gates at the same level were swept from 0~V to $-0.8$~V. All the 32 MUX channels are completely pinched off when the gate voltage reaches -0.66V. Inset shows the pinch off characteristics of MUX channels that are covered by 500nm polyimide. It was measured with the same input signal of -0.4V, however only one addressing gate was used to pinch off channels. Half of channels that are not covered by polyimide are pinched off at -0.66V, while the other half are pinched off at $-7.2$~V. (b) The leakage currents of all 32 channels, when a pair of gates at each level were set to pinch off channels respectively. The leakage currents are plotted with 10~pA offset for clarity. (c) Pinch off characteristics of MUX channels when the input signal was set to $-0.5$~V and $-0.6$~V respectively. Evidently, more negative gate voltage is required to pinch off channels. (d) Multiplexed measurement of a nm-size graphene device, where various positive voltages are applied to inactive gates. As can be seen, the current saturation in MUX channel can obscure the intrinsic characteristics of the qDUT and lead to measurement artefact, if the MUX is not operated properly.
\label{MUX-GNR}}
\end{figure*}

Figure \ref{MUX-GNR}a shows a representative channel pinch off characteristics of a MUX. It was measured by sweeping the voltage applied to a pair of addressing gates at a same level simultaneously, while the MUX input was kept at $V_{sd} = -0.4$~V. All the channels are pinched off when the voltage applied to the addressing gates reaches $V_{g} = -0.66$~V, which is equivalent to a threshold voltage  $V_{T}$ of $-0.26$~V. On the other hand, the inset shows the pinch off characteristics when negative voltage is only applied to a single addressing gate. Half of the channels without polyimide are pinched off at $V_{g} =- 0.66$~V (not shown), while all other channels covered by polyimide are not pinched off until  $V_{g} = - 7.2$~V. There is indeed a window of gate voltage that the channels can be selectively pinched off. Figure \ref{MUX-GNR}b shows the leakage current of 32 pinched off channels of a MUX, when two addressing gates of each level were set to $-0.66$~V and the input signal was swept from $-0.4$~V to 0.4~V. The leakage currents are plotted with 10~pA offset for clarity. As can be seen, the total leakage current of 32 channels is less than 1~pA. By fitting the total leakage current, we infer the off-resistance of each channel is around 30~T$\Omega$. For a MUX with 256 channels, the effective resistance of all closed channels is still above 100~G$\Omega$ and will allow us to measure qDUTs up to $\sim$ 10~G$\Omega$ resistance. From the leakage current, we also learn indirectly that the noise added by MUX circuitry is negligible, since no significant change is observed as compared with the noise floor set by electronics.

Figure \ref{MUX-GNR}c shows the pinch off characteristics of the same MUX, when the input signal was set to $-0.5$~V and $-0.6$~V respectively. It is evident that more negative gate voltages are required to pinch off channels. Generally, to ensure active addressing gates pinch off all unselected channels completely, input signal $V_{in}$ is limited to $V_{g} - V_{T}$ at the negative side, where $V_{g}$ is the gate voltage applied and $V_{T}$ is the threshold voltage. Therefore, to extend the negative limit of input signal $V_{in}$ is essentially to extend the negative limit of gate voltage $V_{g}$ that can be applied. As each addressing gate controls multiple channels, the potential difference between channel and addressing gate also varies. It can be found out that the smallest potential difference is $V_{g} - V_{in}$ and the biggest potential difference is $V_{g}$ (see Figure \ref{Breakdown}a as an example). It is the channels that have the biggest potential difference with respect to the active addressing gate setting the negative limit of $V_{g}$. By incorporating 20~nm Al\textsubscript{2}O\textsubscript{3} layer, the negative limit of $V_{g}$ is substantially increased, which would otherwise be set by the breakdown voltage of reversely biased Schottky junction formed between addressing gates and channels. From the measurement to obtain the pinch off voltage of channels that are covered by polyimide (Figure \ref{MUX-GNR}a inset), we learn that the voltage applied to addressing gate can be at least as large as $-10$~V, since channels without polyimide are also being gated up to $-10$~V in that measurement.

On the other hand, the input signal range at positive side is limited by inactive addressing gates. Figure \ref{MUX-GNR}d shows the multiplexed measurement of 32 $\mu$m-size graphene ribbons, where active addressing gates were set to $-0.66$~V and inactive addressing gates were left grounded. The bias voltage $V_{sd}$ was supplied to the MUX input and swept over a range from $-0.4$~V to $0.4$~V. All the graphene ribbons share a common drain, from which the current was measured (Figure \ref{graphene-qDUT}). As can be seen, the current measured flowing through each MUX channel is saturated at high positive bias, which is not an intrinsic characteristic of $\mu$m-size graphene ribbons but rather originates from the MUX channel. As the input signal (bias voltage) increases at the positive side, the inactive addressing gates (kept grounded) become effectively more negative with respect to the open MUX channel, such that the 2DEG is gradually depleted (Figure \ref{MUX_Operation}b) and any further increase of the bias voltage is dropped across the depleted region. The input signal range at positive side is thus limited. Each channel saturates at slightly different voltage, which is due to the variation in the local carrier density of 2DEG and the gate capacitance combined. To prevent the MUX channel from going into saturation and extend the positive limit of the input signal, a positive voltage can be applied to inactive addressing gates. Figure \ref{MUX-GNR}e shows the multiplexed measurement of the same 32 $\mu$m-size graphene ribbons, but the inactive addressing gates were set to 0.4~V. As can be seen, the currents are no longer saturated and show linear current-voltage relationship as expected from $\mu$m-size graphene ribbons. The positive voltage that can be applied to the inactive addressing gates itself is limited. Without the dielectric layer, it is set by the turn-on voltage of the Schottky junction ($\sim$ 500~mV), beyond which large gate leakage current to unselected channels will be generated (see Figure \ref{Breakdown}b as an example). By adding 20~nm Al\textsubscript{2}O\textsubscript{3}, the gate leakage can be effectively suppressed until dielectric breakdown occurs. Although we didn't test the breakdown voltage at the positive side, we can safely estimate it to exceed 10~V based on two reasons. First, unlike Schottky junction, metal dielectric interface has little dependence on voltage polarity, so we expect it to be comparable to the negative side. Second, based on the typical breakdown electric field reported for ALD grown Al\textsubscript{2}O\textsubscript{3} on GaAs substrate \citep{Wu2007}, dielectric breakdown voltage for 20~nm Al\textsubscript{2}O\textsubscript{3} can be as large as 20~V at room temperature and can be a few times higher at cryogenic temperature. 

The dotted line in Figure \ref{MUX-GNR}e shows the current trace of a open MUX channel measured separately without any qDUT connected, which is fitted to have an on-resistance of 862~$\Omega$. It is much smaller as compared with the resistance of a typical qDUT, which is at least a few tens of k$\Omega$. Figure \ref{MUX-GNR}f shows the current trace of a graphene qDUT measured with different positive voltages applied to inactive addressing gates. As can be seen, when no voltage is applied, the current trace shows a plateau above 0.2~V. As the voltage increases to 0.2~V, the onset of saturation is shifted to higher voltage. When the accumulation voltage is set to 0.4~V, the plateau disappears completely. This current plateau is due to 2DEG saturation and can be wrongly attributed to resonant transport behaviour, if the MUX is not operated properly.

\section{Measuring graphene nanogaps}
We further demonstrate the MUX setup by performing quantum transport measurement on nm-size graphene nano-gaps. Graphene nanogap has been developed to contact single molecule and to study quantum transport across the single molecular junction \citep{Lau2014}. As compared with conventional approaches based on scanning probe or metal break junction, it allows easy integration of a backgate to tune the molecular orbitals. Moreover, the reduced electric field screening of graphene allows multiple molecular orbitals to be accessed for transport study. Most importantly, its planar device structure holds the prospect to scale up and build functional molecular circuit. Given all these advantages, it does have a disadvantage that it is difficult to collect sufficient device statistics, thereby to provide feedback for optimisation. This problem has to be overcome before building any functional molecular circuit becomes possible. Take a work done previously in our group for example \citep{Limburg2018}, which assessed the effectiveness of different anchor groups to form molecular junctions. In total it took three people two years of experiments to collect statistics on 1000 devices. In comparison, the multiplexed measurement of 128 devices only took 5 days to finish. Same amount of statistics will only need 40 days, which is roughly 20 times speed up.

\begin{figure*}
\includegraphics[width=0.98 \linewidth]{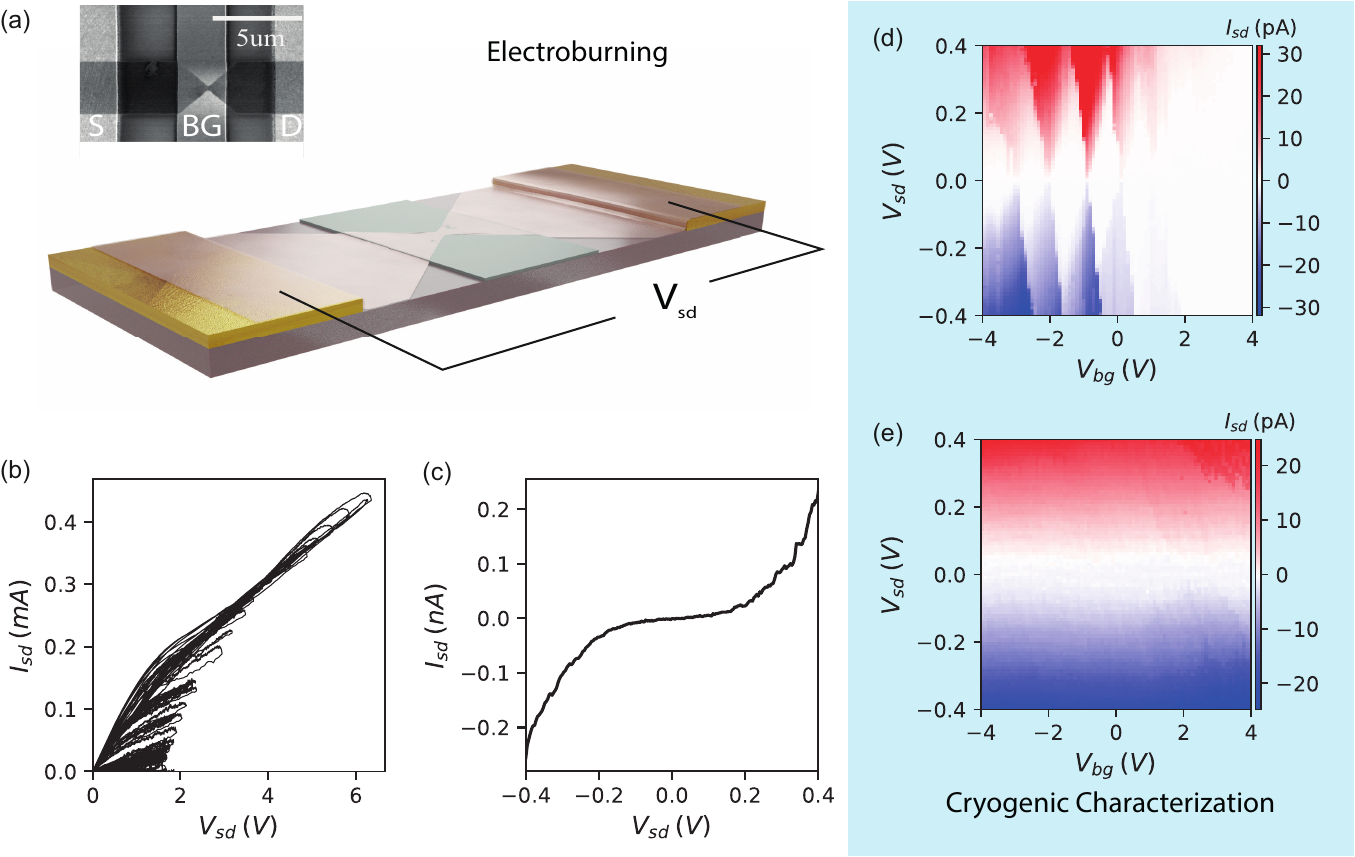}
\caption{(a) Schematics of a graphene nano-constriction, which is used to fabricate a graphene nano-gap by electroburning process. Inset shows the SEM of a graphene nano-constriction. (b) Electroburning curves of the graphene nano-constriction. At each burning cycle, the bias voltage $V_{sd}$ is ramped up at rate of 5V/s to break the graphene, upon the deteciton of a sudden current drop, the voltage is ramped down at a rate of 50V/s. This controlled breakdown of graphene nano-constriction leads to the formation of nm-size graphene nanogap. (c) From the tunneling curve at the end of electroburning, we fit it with Simmons model and infer the gap size to be 1.2~nm. (d) Stability diagram of a graphene nanogap measured at 2.9~K in puck tester. It shows resonant tunneling features, which are indistinguishable from intrinsic charge transport features across molecular junctions, such that this nanogap is not suitable for contacting molecules. (e) In contrast, stability diagram of a different graphene nanogap, which does not show any particular features and is suitable for contacting molecules.
\label{eburn}}
\end{figure*}

Graphene nano-gaps used here were fabricated using the electroburning process\citep{Lau2014}. A difference in device structure is that a local back gate is used instead (Figure \ref{eburn}a). Additionally, local back gate and drain are shared among the graphene devices {Figure \ref{graphene-qDUT}}, such that only one multiplexed interconnect instead of three is needed for each device to perform transport measurement. Figure \ref{eburn}b shows a typical electroburning process. At each burning cycle, the bias voltage, applied between the source(S) and drain (D) across the graphene nano-constriction, is ramped up at a rate of 5V/s. Upon the sudden decrease of the current resulting from the breakdown of graphene constriction, it is ramped down at a faster rate of 50V/s. By repeated 
controlled breakdown, a nanogap forms in the end and shows sub-nA tunneling current (Figure \ref{eburn}c). By fitting into the Simmons model, it can be inferred that the gap size is around 1.2~nm.

Not all the graphene nano-gaps are suitable for contacting molecules, significant percentages of nanogaps have carbon debris in between. At cryogenic temperature it may behave like a quantum dot and show resonant transport features that are indistinguishable from molecule and an example is as shown Figure \ref{eburn}d. If a nanogap like that is used for contacting molecules, any features appear after molecule deposition cannot be simply attributed to the molecule, therefore should be screened before molecule deposition. In comparison, Figure \ref{eburn}e shows a different nanogap, which does not show any Coulomb Diamond alike features and is suitable for contacting molecules.

\begin{figure*}
\includegraphics[width=0.98 \linewidth]{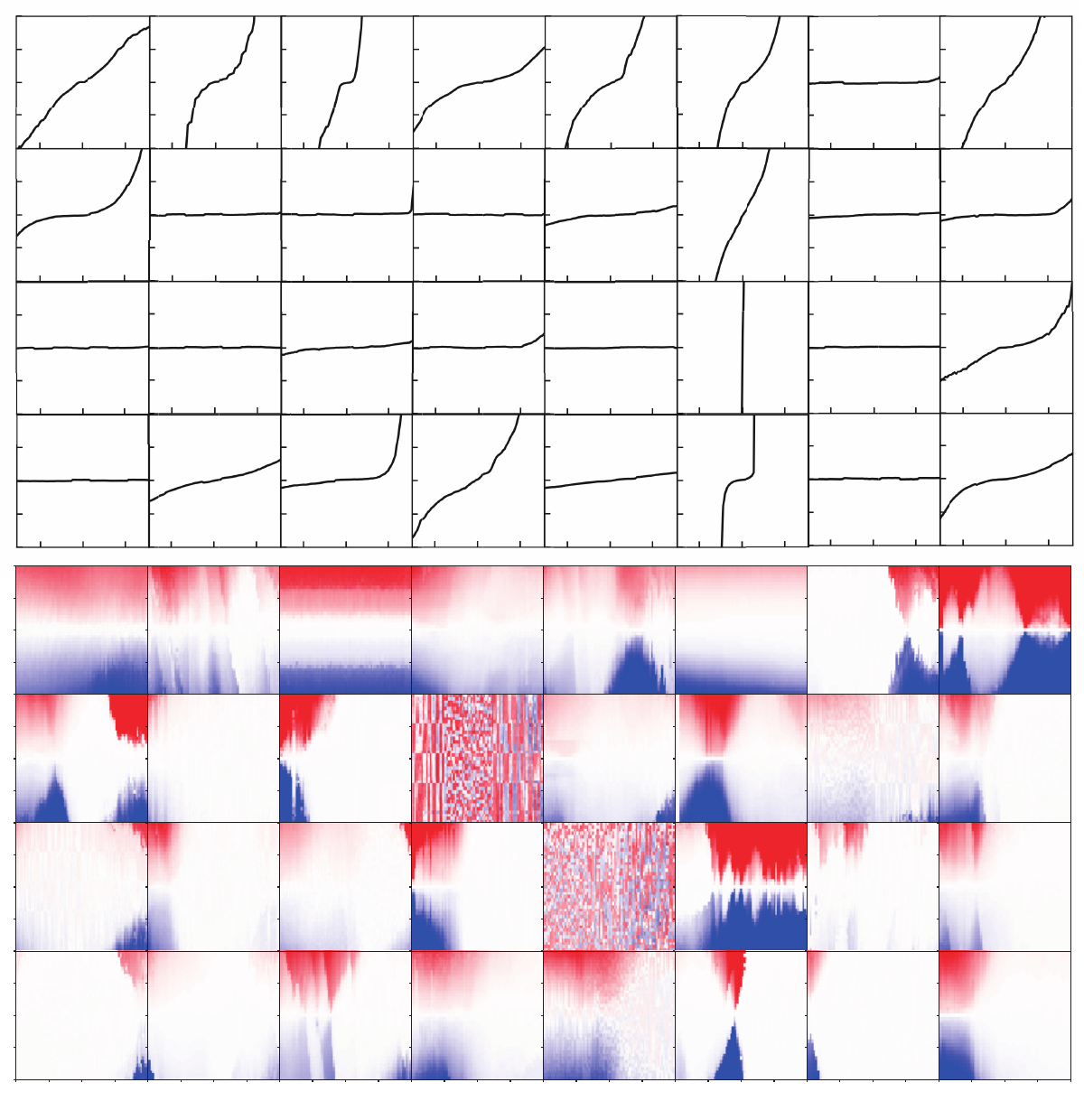}
\caption{Top: IV curves of 32 graphene nanogaps measured at the end of electroburning process at room temperature. They were measured with a bias voltage $V_{sd}$ range from $-0.4$~V to $0.4$~V and zero back gate voltage $V_{bg} =0$~V. All the curves are plotted in the same scale, the vertical axis is between -100pA to 100~pA. Bottom: Stability diagram of the same 32 graphene nanogaps measured at 2.9~K in puck tester. The measurement was performed with a bias voltage $V_{sd}$ range from $-0.4$~V to 0.4~V and a back gate voltage $V_{bg}$ range from $-4$~V to 4~V. The 32 stability diagrams are plotted in a same colorbar, where different shades of blue corresponds to the range from $-500$~pA (blue) to 0~pA (white) and red corresponds to 500~pA (red) to 0~pA (white).
\label{MUX_stab}}
\end{figure*}

Usually we perform electroburning at multiple back gate (BG) voltages, which makes it more likely to form a featureless gap like Figure \ref{eburn}e. However, for the purpose of demonstrating MUX setup, i.e. to easily distinguish different gaps being measured, all the graphene nano-gaps measured were fabricated with gate floated. We measured 128 graphene nano-gaps in total, 32 of them are shown in Figure \ref{MUX_stab} and all other 96 of them are shown in Figure \ref{stab_96}. In Figure \ref{MUX_stab} top panel shows the tunneling curves measured at the end of electroburing process at room temperature and bottom panel shows the stability diagrams measured at 2.9K in puck tester (Figure \ref{setup}). The tunnelling curves are plotted in the same scale, where the bias range is from $-0.4$~V to 0.4~V and the current range is from $-500$~pA to 500~pA. The stability diagram is plotted with the bias range from $-0.4$~V to $0.4$~V and a back gate voltage range from $-4$~V to $4$~V.

\section{Conclusion and Outlook}
The prototype setup demonstrated in this paper can supply in total 128 interconnects at cryogenic temperature, which uses only 14 out of 44 existing wires from room temperature. As can be seen from Figure \ref{setup}b, the 2 MUX chips (4 MUXs) are still sparsely placed on the PCB. Even with our very primitive 2 layer PCB (minimum track size 0.2~mm, drill size 0.3~mm), we can easily fit 1 more MUX chip (as shown in Figure \ref{MUX_192}). This will give us in total 192 interconnects, of which 10 shared control lines and 6 input signal lines are used. If multiple layer PCB is used and all the tracks placed in inner layers, there will be space for mounting up to 9 MUX dies (18 MUXs). Another aspect that can easily be improved is the PCB track size and the footprint of board to board connector (Figure \ref{setup}c). If the size is reduced by half, we can add one more level to the MUX and fit it roughly into the same space. With 9 MUX dies mounted, they can supply 1152 interconnects in total with only 30 (12+18) wires used.

To conclude, we have successfully demonstrated a MUX setup to increase the throughput of electrical characterisation at cryogenic temperature. We have also assessed various limits that exist in MUX operation. By implementing the MUX circuitry off-chip, the qDUT chips can be easily exchanged. Different qDUT PCBs are designed to meet the wiring requirements of different qDUTs. Moreover, we have incorporated a shared control signal arrangement at board level to support the MUX operation. Such implementation will not only provide an efficient to scale up the total number of interconnects, but also will allow us to easily control multiple MUXs simultaneously to interface with qDUTs requiring multiple signal lines to operate.

\section{Acknowlegement}
JAM was supported through the UKRI Future Leaders Fellowship, Grant No. MR/S032541/1, with in-kind support from the Royal Academy of Engineering.

\clearpage
\bibliography{MUX}

\section{Supplemantary Material}
\renewcommand{\thefigure}{S\arabic{figure}}
\setcounter{figure}{0}

\begin{figure*}
\includegraphics[width=0.8 \linewidth]{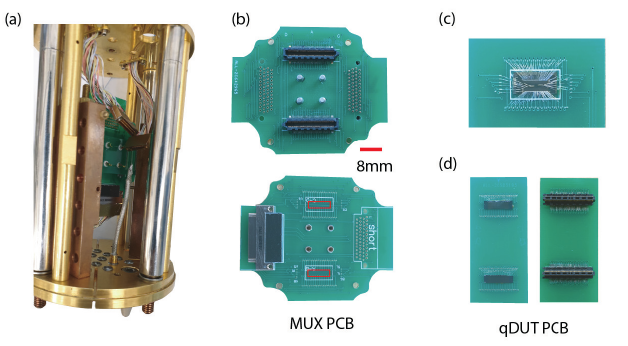}
\caption{(a) Measurement setup used in this paper. The puck is compatbile with both 2.9K puck tester and mK dilution fridge of Oxford Instrument. (b) MUX PCB: the two red rectangles are for mouting MUX chip. With this PCB, we can install 4 MUXs in total (2 MUXs in each chip). 2 layer PCBs are used here and most of board area is occupied by the PCB tracks. If multiple layer PCBs are used and all the tracks put into the inner layer, there is enough space to accomodate up to 9 MUX chips. The board to board connector used here is of 0.4mm pitch size and each contsist of 80 pins. (c) The MUX PCBs are made with primitive technology, of which the minimal track width is 0.2mm (equivalent to a pitch size of 0.4mm). If PCBs are made with a track width of 0.1mm, MUXs with 64 output channles roughly fit into the same space . (d) qDUT PCB used in this setup.
\label{setup}}
\end{figure*}

\begin{figure*}
\includegraphics[width=0.8 \linewidth]{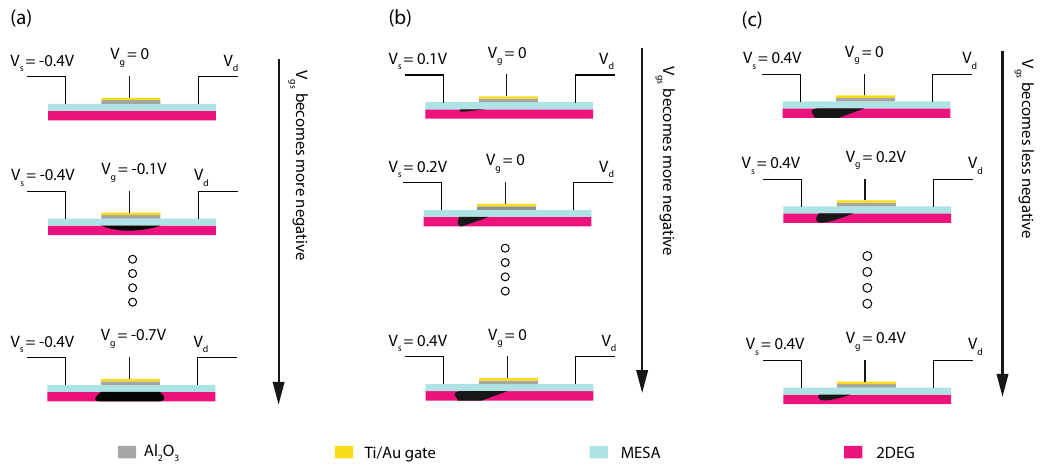}
\caption{(a) Schematics showing the channel pinch off process. As the gate voltage becomes more negative with repsect to the MUX channel, the 2DEG is gradually depleted. (b) Schematis showing the 2DEG saturation process. As the input signal increases and 2DEG becomes more positive with respect to addressing gate, which acts effectively as if a negative gate voltage was applied, thus the 2DEG is depleted closer to the input signal side. The current saturation arises a result, since any further increase of input signal is dropped over the depleted region. (c) A positive voltage applied to the gate can reduce the potential difference between addressing gate and channel, so to prevent the 2DEG going to saturation. The signal range at the positive side can thus be extended.
\label{MUX_Operation}}
\end{figure*}

\begin{figure*}
\includegraphics[width=0.8 \linewidth]{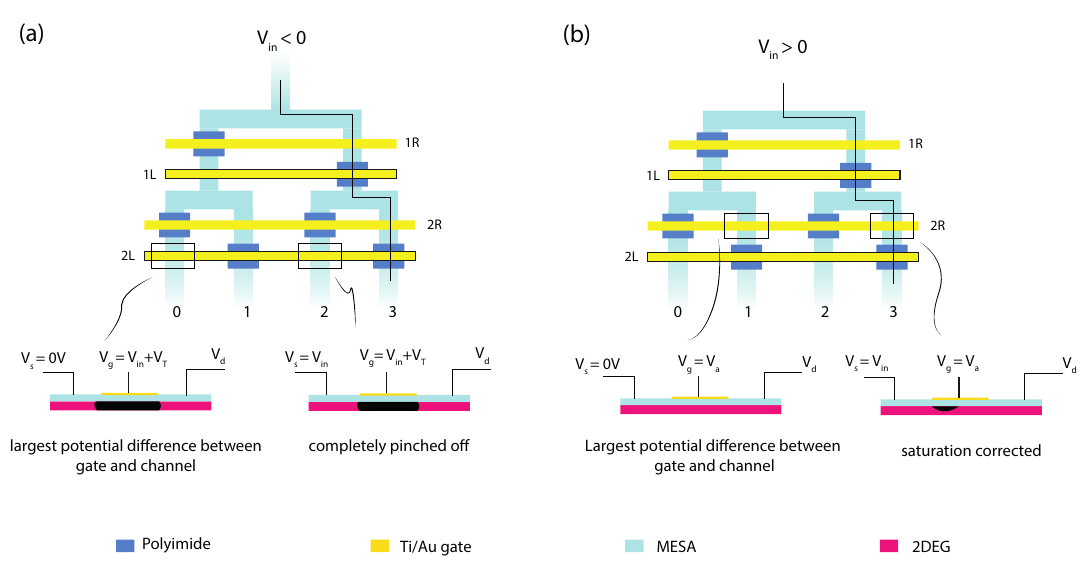}
\caption{Schematics of a 2-level base-2 MUX, where addressing gate 1L and 2L are activated to select channel 3. (a) The negative limit of input signal is set by the negative voltage applied to addressing gate 1L and 2L. The gate voltage itself is limited by the channel that has the biggest potential difference with respect to active addressing gate. Here it is between channel 0 and gate 2L, of which the potential difference is V\textsubscript{in}. (b) The positive limit of input signal is set by the inactive addressing gates. To prevent the 2DEG going to saturation, a positive voltage can be applied to 1R and 2R to extend the input signal range at positive side. However, the positive voltage that can be applied is also limited by the channel that has the biggest potential difference with respect t inactive addressing gate. Here it is between channel 1 and addressing gate 2R.
\label{Breakdown}}
\end{figure*}

\begin{figure*}
\includegraphics[width=0.8 \linewidth]{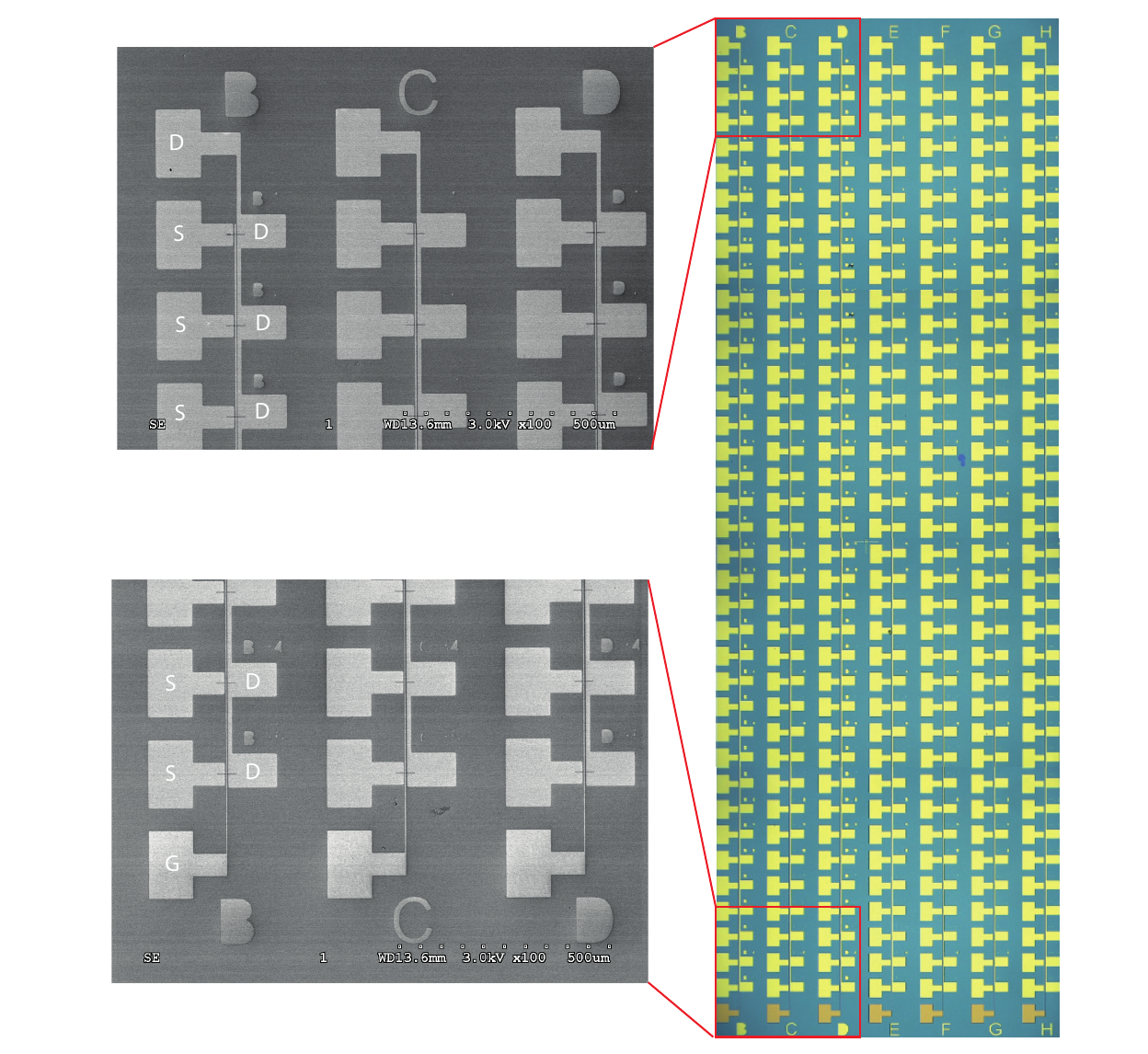}
\caption{Silicon substrate with electrodes for fabricating graphene used in this paper. Both $\sim$ $\mu$m size graphene ribbons and 100nm graphene nanoconstrictions are made with this device structure, where all the devices of a column share a common drain and a common local back gate. The local back gate is covered is 10nm HfO\textsubscript{2} grown by ALD.
\label{graphene-qDUT}}
\end{figure*}

\begin{figure*}
\includegraphics[width=0.8 \linewidth]{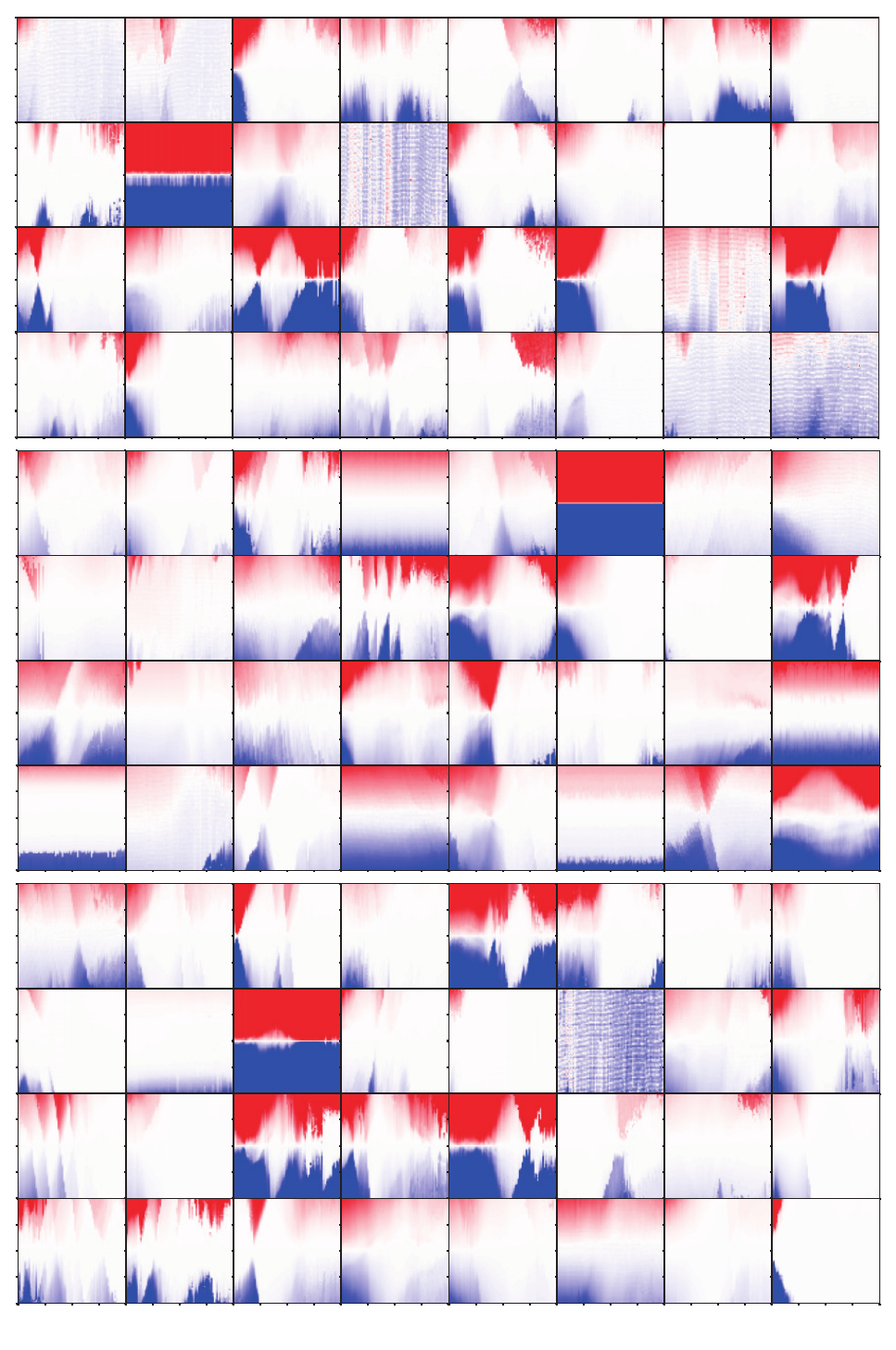}
\caption{Stability diagrams of all other 96 graphene nanogaps which were measured in single cooling down. The measurement was performed at 2.9K in puck tester. The measurement was performed with a bias voltage V\textsubscript{sd} range from -0.4V to 0.4V and a back gate voltage V\textsubscript{bg} range from -4V to 4V. The 32 stability diagrams are plotted in a same colorbar, where different shades of blue corresponds to the range from -500 pA (blue) to 0 pA (white) and red corresponds to 500 pA (red) to 0 pA (white).
\label{stab_96}}
\end{figure*}

\begin{figure*}
\includegraphics[width=0.8 \linewidth]{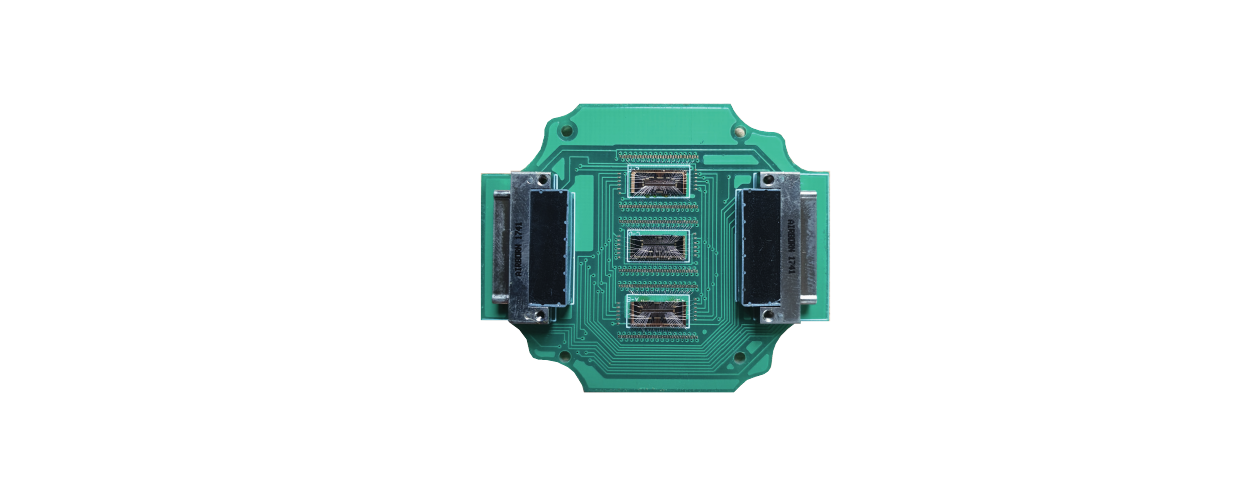}
\caption{A different MUX PCB that can mount 3 MUX chips. In total 192 interconnects can be controlled using 16 wires from room temperature.
\label{MUX_192}}
\end{figure*}

\end{document}